\documentclass{article}
\usepackage[utf8]{inputenc}
\usepackage[T1]{fontenc}
\usepackage{amsmath,fullpage,enumerate}
\usepackage{algpseudocode, algorithm}
\usepackage{booktabs}
\usepackage{microtype}
\usepackage{xcolor}
\usepackage{amsmath}
\usepackage{bm}
\usepackage{graphicx}
\usepackage{makecell}
\usepackage{overpic}
\usepackage{tabularx}
\usepackage{subcaption}
\usepackage{multirow}
\usepackage{array}
\usepackage{gensymb}
\usepackage{authblk}
\usepackage{cite}
\usepackage{url}
\usepackage[normalem]{ulem}
\newcolumntype{L}[1]{>{\raggedright\let\newline\\\arraybackslash\hspace{0pt}}m{#1}}
\newcolumntype{C}[1]{>{\centering\let\newline\\\arraybackslash\hspace{0pt}}m{#1}}
\newcolumntype{R}[1]{>{\raggedleft\let\newline\\\arraybackslash\hspace{0pt}}m{#1}}

\DeclareMathOperator*{\argmin}{arg\,min}
\newcommand{\vect}[1]{\boldsymbol{\mathbf{#1}}}

\floatname{algorithm}{Procedure}

\usepackage{amssymb}

\renewcommand{\cal}{\mathcal}

\newcommand{\highlight}[1]{{\color{black} #1}}

\title{A Variable Neighbourhood Descent Heuristic for Conformational Search Using a Quantum Annealer}

\author[1]{D. J. J. Marchand}
\author[1]{M. Noori}
\author[1]{A. Roberts}
\author[1]{G. Rosenberg}
\author[1]{B. Woods}
\author[1]{U. Yildiz} 
\author[2]{M.~Coons}
\author[2]{D. Devore}
\author[2]{P. Margl}
\affil[1]{1QB Information Technologies (1QBit), 458-550 Burrard Street, Vancouver, BC, Canada, V6C 2B5}
\affil[2]{The Dow Chemical Company, Core R\&D, 1776 Building, Midland, MI, United States, 48674}
\affil[{ }]{\small{Names listed by affiliation and in alphabetical order.}}

\begin{document}
\maketitle

\begin{abstract}
Discovering the low-energy conformations of a molecule is of great interest to computational chemists, with  applications in {\em in silico} materials design and drug discovery. In this paper, we propose a variable neighbourhood search heuristic for the conformational search problem. Using the structure of a molecule, neighbourhoods are chosen to allow for the efficient use of a binary quadratic optimizer for conformational search. The method is flexible with respect to the choice of molecular force field and the number of discretization levels in the search space, and can be further generalized to take advantage of higher-order binary polynomial optimizers. It is well-suited for the use of devices such as quantum annealers. After carefully defining neighbourhoods, the method easily adapts to the size and topology of these devices, allowing for seamless scaling alongside their future improvements.
\end{abstract}

\flushbottom
\maketitle
\thispagestyle{empty}

\section{Introduction}\label{sec: intro}
The study of molecular structures is foundational to attaining an understanding of chemical processes. Chemical behaviour is determined in large part by the arrangement of atoms within participating molecules as a chemical process unfolds. A widely used approach for studying aspects of molecular structure is to limit the molecular degrees of freedom to torsions only, considering bond lengths and bond angles to be fixed at some  values. By doing so, a given molecule with a specific connectivity of its constituent atoms may take on a variety of three-dimensional spatial arrangements, known as  \emph{conformational isomers}, or simply \emph{conformations}. Whereas the internal bonds of distinct conformations are the same, the rotation angles around rotatable bonds of the molecule are what distinguish one conformation from another. 

In medicinal chemistry, conformational analysis often involves identifying bioactive conformations of ligand molecules~\cite{perola2004conformational,kirchmair2005comparative}. Protein folding is one illustration of function determined by conformation that is both very important to the fields of medicine and biochemistry and extremely difficult to compute, making it a ``grand challenge'' for science~\cite{dill2012protein}. In a broader context, conformational analysis is a topic of utmost importance in chemical and materials research. For instance, the conformational behaviour of polymers is key to determining  crystallinity, shape, and entanglement of individual chains~\cite{boyd1996science} that in turn affect macroscopic materials' properties such as elasticity, strength, or toughness. 

Importantly, the geometrical differences between conformations result in different values for the molecular potential energy~\cite{miao2016unconstrained,copeland2011conformational,izgorodina2007energy}, which is a key factor for molecular stability and reactivity. To this end, the problem of finding the conformations associated with the local minima of the potential energy surface (PES) of the molecule, referred to as the \emph{conformational search problem}, has been of long-standing interest. The local minima of the PES are often called~\emph{conformers}~\cite{hawkins2017conformation}, and the conformer with the lowest energy is referred to as the \emph{global minimum-energy conformer}. 

Over time, several practical approaches to the conformational search problem have been developed and critically reviewed~\cite{hawkins2017conformation, leach1991survey, pd2014conformational, loferer2007analyzing}. For small molecules, one may be able to deterministically solve the conformational search problem via techniques such as branch and bound~\cite{bruccoleri1987prediction, goodman1991unbounded, christen2008searching}. However, this is impractical for even moderately sized molecules, as the size of the conformational search space grows exponentially with the number of rotatable bonds~\cite{goedecker2004minima}. Such increasing difficulty makes the conformational search problem computationally intractable for many of the molecules that have real-world applications and an attractive target for exploration with novel computational technologies and techniques.  

To address the computational complexity presented by larger molecules, many metaheuristic approaches have been studied. Examples of such approaches include genetic algorithms \cite{vainio2007generating,supady2015first}, conformational space annealing \cite{lee1997new,heo2017protein}, tabu search \cite{morales2000parallel,stepanenko2009tabu}, molecular dynamics (MD) \cite{li1998conformational,doshi2015towards}, and basin/funnel hopping \cite{gehrke2009assessing}. Variations of the Monte Carlo (MC) method have also been widely used~\cite{wilson1991applications,sperandio2009med} as a less computationally expensive alternative to MD \cite{hawkins2017conformation}. In addition, parallel tempering (PT), also known as replica exchange, \cite{thachuk2007replica,rhee2003multiplexed} can be applied to both MC and MD to further improve their sampling performance of the conformational search space. 

A comparatively recent approach to address the growing computational complexity of optimization problems relies on the putative future advantage of specialized hardware like quantum annealers to solve binary quadratic optimization problems (see Supplementary Information for more details). One challenge lies in reformulating the optimization problems, a task that often requires approximations or simplifications. Our motivation was to develop such a formulation for the conformational search problem that avoids drastic  compromises while producing good conformers.  

For this purpose, we propose an iterative heuristic method for the conformational search problem based on \emph{variable neighbourhood descent} (VND). In each iteration of the method, we use the molecular structure to choose specialized conformational neighbourhoods that can be minimized efficiently. More specifically, using the structural graph of a given molecule, subsets of rotatable bonds are selected at each iteration. Fixing the values of other torsion angles, the problem of minimizing the molecular energy with respect to the selected torsion angles is then formulated as a binary program with an objective function that is a polynomial of a chosen degree. This allows the method to be adapted to the specifics of the optimizer by limiting the degree of the binary program. The values of the selected torsion angles are then set to the solution of this binary program before starting each subsequent iteration during which a new subset of rotatable bonds are optimized. The process continues until some stopping criteria are met. 

Although the method can be readily extended to any chosen degree, we assume in this paper that a binary quadratic program is desirable as it is well-suited for optimization using  quantum annealing~\cite{santoro2002theory,morita2008mathematical}. Furthermore, by changing the parameters of the neighbourhood selection procedure, the method can be easily adapted to the size of the conformational search problem in terms of the number of rotatable bonds, as well as the size and connectivity of available quantum annealers. The flexibility of solving any conformational search problem using currently available quantum annealers, without imposing restrictions on the granularity of the conformational space, differentiates our work from a previous study on protein folding using quantum annealing~\cite{perdomo2012finding}. 

We evaluate the performance of our proposed algorithm over three families of molecules relevant to industry, using an algorithm that returns an exact optimal solution and the D-Wave 2000Q quantum annealer~\cite{Dwave_2000Q, van2007adiabatic,johnson2011quantum}. The latter provides an assessment using the latest available hardware at the time of writing of this work, whereas the former can be seen as a limiting ideal case. For each molecule, we compare the lowest-energy conformations found by our algorithm with those found by both parallel tempering MC (PTMC)~\cite{swendsen1986replica,thachuk2007replica} and a simple local search method.

\section{Preliminaries} \label{sec: preliminaries}
We give some preliminaries before presenting the details of our proposed conformational search method. 

\subsection{Problem Definition}
As discussed above, we consider the conformational search problem as a special case of molecular structure analysis, where the structure is kept fixed except for rotations around selected bonds. Each of these torsional degrees of freedoms we hereafter refer to as a \emph{torsion} for simplicity. We denote the $i$-th rotatable bond by $T_i$ and assign its rotational angle a variable $t_i$, with $i$ representing the torsion index. It is convenient to identify a conformation of a molecule with $M$ torsions by a \emph{torsion vector} $\vect{t} = [t_1,\, \ldots,\, t_M]$. Without loss of generality, we assume $t_i \in [0,2\pi)$, for all $i$, knowing that the method remains unchanged if each torsion has its own range chosen based on prior knowledge, experimental data, or known symmetries. For simplicity, let us assume all torsion angle values are chosen from the same set of $d$ values $\Theta = \{ \theta_1,\, \ldots,\, \theta_d \}$. The theoretical precision of this discretization scheme increases with $d$, while the size of the search space $d^M$ grows exponentially with the number of torsions. 

\begin{figure*}[!t]
\centering
\includegraphics[width=\linewidth, clip]{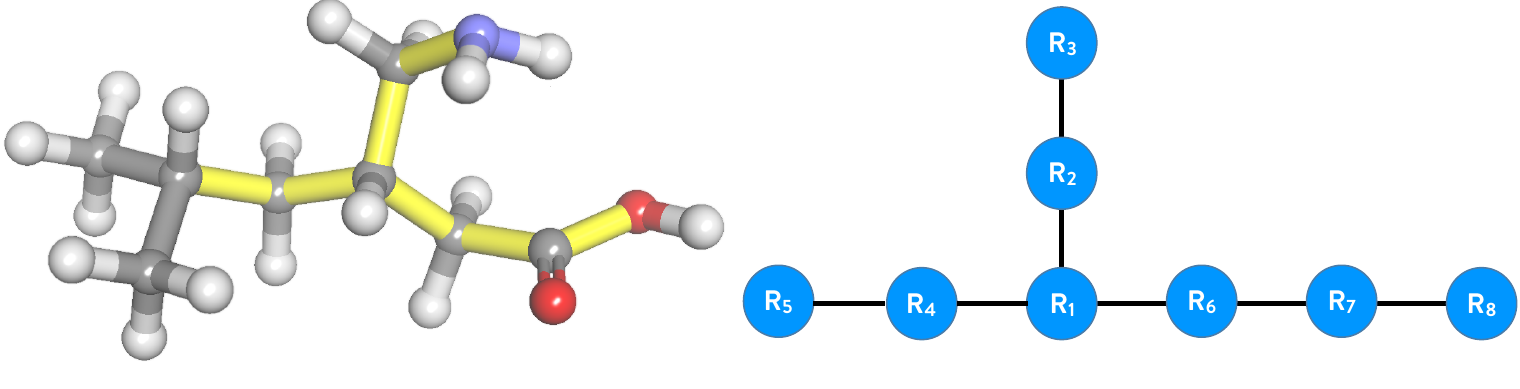}
\caption{Illustration of the rigid-body graph for a simple organic molecule (pregabalin). The molecular structure, with seven rotatable bonds highlighted in yellow, is shown on the left. The rotatable bonds connected to the two methyl groups have been discarded for visual clarity. The associated rigid-body graph is depicted on the right.}
\label{Fig: rb graph example}
\end{figure*}

Although it is natural to describe a molecule using a molecular graph, where the atoms and their bonds are represented by vertices and edges, respectively, we find it helpful to use the torsions to partition the molecule into $M+1$ subsets called \emph{rigid bodies}. The partitioning is performed such that all atoms within a rigid body are interconnected through non-torsion bonds. As a result, the relative positions of the atoms within a rigid body, denoted by $R_a$, remain invariant under rotation and are therefore independent of $\vect{t}$. This simplified representation of the molecule is now easily described by a \emph{rigid-body graph} $G=(\mathcal{R}, \mathcal{T})$, where $\mathcal{R}$ is the set of $M+1$ vertices and $\mathcal{T}$ is the set of $M$ edges. In $G$, each vertex represents a rigid body and each edge represents a rotatable bond. Two vertices are connected by an edge if their associated rigid bodies are connected by the rotatable bond that the edge represents. We will therefore use $T_i$ to refer to both torsion $i$ and its associated edge in the rigid-body graph. We further assume that each torsion is free to rotate independently of others, thus restricting the presence of ring systems or other cycles in the molecular graph to  individual rigid bodies. Under this assumption, the rigid-body graph has no cycles and is a tree. An example of a simple molecule and its rigid-body graph is shown in Fig.~\ref{Fig: rb graph example}.

The search space of the conformational search problem is a hypersurface described by an energy model (function) $U: \Theta^M \rightarrow \mathbb{R}$. For a given $\vect{t}$, $U(\vect{t})$ is the molecular energy consisting of the sum of all interatomic potentials in the molecule (e.g., van der Waals, torsional, and bending), which are dependent on the relative coordinates of the atoms. Various energy models or effective \emph{force fields} can be used for our purpose, such as the widely used ``Universal force field'' (UFF)~\cite{Rappe1992}. The conformational search problem, with the objective of finding the global minimum-energy conformer of a given molecule, can then be formulated as
\begin{align} \label{eq: general optimization}
\displaystyle{\min_{\vect{t}}} \quad & U( \vect{t}) \\ \nonumber
\text{s.t.} \quad  & \vect{t} \in \Theta^M.
\end{align}

Upon changing the torsion angles of the molecule, some of the interatomic potential contributions will remain unchanged, while other contributions will change depending on the torsion angle vector $\vect{t}$. To be more specific, let us denote the values of the torsion angles on the path connecting $R_a$ to $R_b$ on the rigid-body graph by a vector $\vect{t}_{ab}$. The length of this path is represented by $m_{ab}$, meaning that $\vect{t}_{ab}$ has $m_{ab}$ elements. The relative positions of the atoms in $R_a$ with respect to the atoms in $R_b$ depend only on the torsions on this path. Now, 
\begin{equation} \label{eq: U(t)}
U({\bf \vect{t}}) =  \sum_{a:R_a \in\mathcal{R}} U_a + \sum_{\substack{a,b:R_a,R_b \in \mathcal{R} \\ a \neq b}}  U_{ab}( \vect{t}_{ab})\,,
\end{equation}
where $U_{a}$ is the sum of the interatomic potentials of the atoms within rigid body $R_a$, which is invariant under rotation, while $U_{ab}:\Theta^{m_{ab}} \rightarrow \mathbb{R}$ such that for a given $\vect{t}_{ab}$, $U_{ab}( \vect{t}_{ab})$ is the sum of the interatomic potentials of all pairs of atoms where one atom is in $R_a$ and the other is in $R_b$. 

\subsection{Binary Optimization Formulation for the Conformational Search Problem} \label{subsec: binary formulation}
In order to use a quantum annealer to solve the conformational search problem, one needs to reformulate it as a quadratic unconstrained binary optimization (QUBO) problem (see the Supplementary Information for more details). To this end, we start by applying a one-hot encoding to the discrete values of the torsion angles, establishing a mapping between the torsion angle vector space and a binary solution space. That is, for each $t_i$, we assign a binary variable $x_{ik}$, $1 \leq k \leq d$, such that 
\begin{equation}
x_{ik} = \begin{cases} 
1 & \text{if } t_i = \theta_k\,; \\
0 & \text{otherwise.} 
\end{cases}
\end{equation}
As a result, $t_i$ can be expressed as
\begin{equation}  \label{eq: one-hot encoding}
t_i = \sum_{k=1}^d \theta_k x_{ik}\,, 
\end{equation}
where we add a constraint
\begin{equation} \label{eq: one-hot constraint}
\sum_{k=1}^d x_{ik} =1
\end{equation}
to ensure $t_i$ takes one and only one value at a time. The constraint \eqref{eq: one-hot constraint} is commonly referred to as a one-hot encoding constraint. Note that after applying the one-hot encoding, any arbitrary function $f(t_i)$ can be written as 
\begin{equation}\label{eq: f(t)}
f(t_i) = \sum_{k=1}^d f(\theta_k) x_{ik}\,.
\end{equation}
Similar to equation \eqref{eq: f(t)}, a binary representation for $U(\vect{t}_{ab})$ can be found. For simplicity of presentation, let us assume that the torsion angles in $\vect{t}_{ab}$ are indexed sequentially from 1 to $m$, that is, $\vect{t}_{ab} = [t_1, t_2, \ldots, t_m]$. Now, 
\begin{equation} \label{eq: U(t_{ab})}
U_{ab} (\vect{t}_{ab}) = \sum_{ k_1=1}^d \cdots \sum_{ k_m=1}^d  U_{ab}(t_1=\theta_{k_1}, \ldots, t_m  = \theta_{k_m} ) \prod_{i=1}^m  x_{i\, k_i}\, ,
\end{equation}
where $U_{ab}(t_1=\theta_{k_1}, \ldots, t_m  = \theta_{k_m} )$ can be pre-evaluated for all possible $d^m$ values of $\vect{t}_{ab}$.

Substituting $U_{ab}(\vect{t}_{ab})$ from equation~\eqref{eq: U(t_{ab})} into equation~\eqref{eq: U(t)} results in a representation of the molecular energy $U(\vect{t})$ in terms of the binary vector $\vect{x}$. We denote this representation of the energy function by $E:  \{ 0,1\}^{Md} \rightarrow \mathbb{R}$ and write the conformational search problem \eqref{eq: general optimization} as
\begin{align} \label{eq: general binary optimization} \nonumber
\displaystyle{\min_{\vect{x}}} \quad & E( \vect{x}) \\ 
\text{s.t.} \quad  & \sum_{k=1}^d x_{ik}=1, \quad \forall i \in \{1,\ldots, M \}\,, \\ \nonumber
& \vect{x} \in \{ 0,1\}^{Md}\,.
\end{align}

To solve the above binary optimization problem using a quantum annealer, one faces three challenges. First, the objective function in  formulation~\eqref{eq: general binary optimization} is not necessarily quadratic as $U_{ab}(\vect{t}_{ab})$ may depend on more than two torsions, that is, $m_{ab} > 2$. Second, it is a constrained binary optimization problem. These two challenges indicate that the problem cannot be solved directly on a quantum annealer. The third challenge is that if $\max \limits_{a,b:R_a, R_b \in\mathcal{R}} m_{ab}$ is not much smaller than $M$, constructing an instance of formulation \eqref{eq: general binary optimization} becomes very computationally expensive due to the pre-evaluation of the coefficients in equation~\eqref{eq: U(t_{ab})}. In the following section, we propose a method that addresses these challenges in order to be able to use a quantum annealer for solving the conformational search problem.

\section{Variable Neighbourhood Descent for the Conformational Search Problem}\label{sec: the algorithm}
Neighbourhood search, or local search (LS), is known to be an effective heuristic algorithm for solving a large number of combinatorial optimization problems. In defining a neighbourhood relation between solutions of a problem, local search begins from an initial solution and iteratively explores the neighbourhood of the current solution for improvement. It has been shown that a solution produced by a local search algorithm will often not be globally optimal, but will be suboptimal with respect to another neighbourhood relation~\cite{glover2006handbook}. 
When multiple neighbourhood relations are considered, the algorithm is often referred to as \emph{variable neighbourhood search}~\cite{mladenovic1997variable}. In the context of conformational search, a solution refers to a vector of torsion angles $\vect{t}$.

Let $N_k$, for $k\in \{1,\ldots, K\}$, denote a finite set of neighbourhood structures and $N_k(\vect{t})$ be the set of all solutions in the $k$-th neighbourhood of $\vect{t}$. Starting from an initial solution and a neighbourhood structure, in each iteration, \emph{variable neighbourhood descent} finds the best solution in the neighbourhood of the current solution.  It then updates the current solution with the best solution found, and changes the neighbourhood structure before proceeding with the next iteration. The VND method is summarized in Fig.~\ref{Algo: VND}.

\begin{figure}[h]
\begin{algorithm}[H]
\begin{algorithmic}
\Require Initial solution $\vect{t}_{\mathrm{init}}$
\State $\vect{t}_{\mathrm{c}} \leftarrow \vect{t}_{\mathrm{init}}$ \Comment{Initialize the current solution}
\State Select an arbitrary $N_k\in \{N_1,\ldots, N_K\}$
\While{termination criteria have not been met}
	\State $\vect{t}' \leftarrow \text{arg}\min_{\vect{t}\in N_k(\vect{t}_{\mathrm{c}})} U(\vect{t})$ \Comment{Find the best neighbour in $N_k(\vect{t}_{\mathrm{c}})$}    
	\If{$U(\vect{t}') < U(\vect{t}_{\mathrm{c}})$}
    	\State $\vect{t}_{\mathrm{c}} \leftarrow \vect{t}'$ \Comment{Update the current solution}
    \EndIf
    \State $N_k\leftarrow $ NeighbourhoodChange() \Comment{Change neighbourhoods}
\EndWhile
\end{algorithmic}
\end{algorithm}
\caption{Variable neighbourhood descent method.}
\label{Algo: VND}
\end{figure}

A simple neighbourhood structure is obtained by considering two solutions as neighbours if and only if they differ by exactly one torsion angle value. Such a VND heuristic is exactly the LS heuristic described above. While computationally inexpensive, the performance of LS can suffer in cases where a decrease in the molecular energy cannot be achieved by changing only a single torsion angle value in an iteration. Our proposed VND method improves upon LS by exploring more-complex neighbourhoods. In the following, we describe the components of the method. 

\subsection{Initial Solution}
The initial solution in Fig.~\ref{Algo: VND} can be selected in a variety of ways. One may simply choose a randomly generated torsion angle vector for the given molecule as the initial solution.  Alternatively, one can use a greedy construction method. Another approach is to start from a known high-quality solution. This applies when using our VND method in conjunction with another conformational search method or by exploiting some prior knowledge about a given molecule.

\subsection{Neighbourhood Structures}
\begin{figure}[H]
\centering
\includegraphics[width=0.7\textwidth, trim={70 210 180 250}, clip]{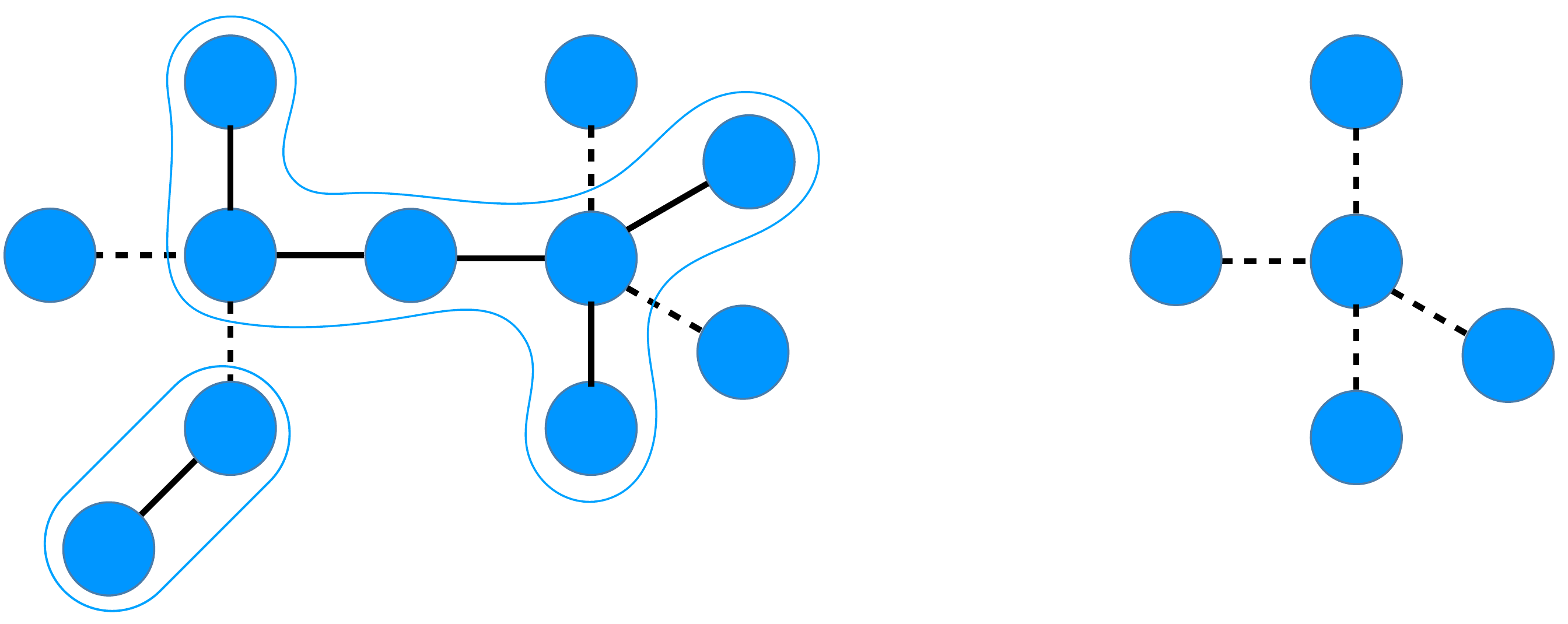}
\caption{Example of a 2-torsion-dependent set of edges in a rigid-body graph (left). The star graph $G'$ (right) results from selecting a 2-torsion-dependent set $\cal{T}'$ (dashed lines) and contracting all edges not in $\cal{T}'$.}
\label{fig: graph contraction}
\end{figure}

We now describe a more powerful neighbourhood structure which, to our knowledge, has not been previously studied.
Let $G=(\cal{R},\cal{T})$ be a rigid-body graph $\cal{T}' \subseteq \cal{T}$, and $G'$ be the graph resulting from contracting all edges in $\cal{T} \setminus \cal{T}'$ (see Fig.~\ref{fig: graph contraction} for an example).  If $G'$ is a star graph (a tree graph with at most one vertex of degree $>1$), then we say that $\cal{T}'$ has the property of \emph{2-torsion dependency}. The motivation for using this terminology is that any two vertices of $G'$ are connected with at most two edges (torsions). We label the maximal 2-torsion-dependent subsets of $\cal{T}$ as $\cal{T}_1,\ldots, \cal{T}_{K}$ and their associated star graphs as $G_1, \ldots, G_{K}$. Then, the neighbourhood structure $N_k$ defines neighbourhoods containing all solutions which  differ only in torsion angle values corresponding to edges in $\cal{T}_k$, for $k=1,\ldots,K$. Solutions in neighbourhoods defined by an arbitrary $N_k$ are also called \emph{neighbours under $N_k$}. 

\begin{figure}[!htb]
\centering
\includegraphics[]{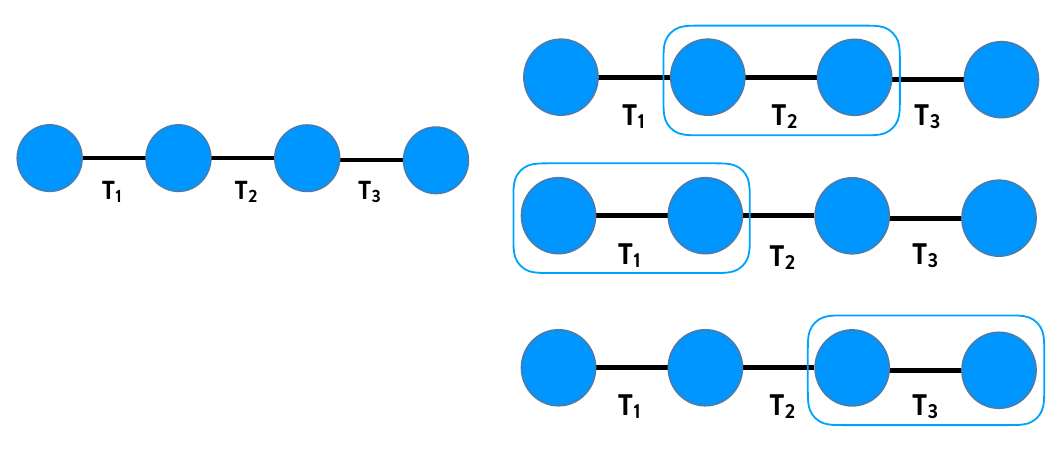} 
\caption{Example of finding different neighbourhood structures of a rigid-body graph. The original rigid-body graph, $G =(\cal{R},\cal{T})$, is shown on the left. All neighbourhood structures defined by the maximal 2-torsion-dependent subsets of $\cal{T}$ are depicted on the right.}
\label{fig:example}
\end{figure}

Neighbourhood structures are illustrated for an example rigid-body graph $G=(\cal{R},\cal{T})$ in Fig.~\ref{fig:example} that represents a molecule with three torsions. The maximal 2-torsion-dependent subsets of the torsions are $\cal{T}_1=\{T_1, T_3\}$, $\cal{T}_2=\{T_2, T_3\}$, and $\cal{T}_3 = \{T_1,T_2\}$. The neighbourhood structure $N_1$ defines neighbourhoods containing all torsion vectors that differ only in torsion angle values for $T_1$ and $T_3$. For example, $\vect{t} = [5\degree,\, 10\degree,\, 20\degree]$ and $\vect{t}' = [0\degree,\, 10\degree,\, 90\degree]$ are neighbours under $N_1$ while $\vect{t} = [5\degree,\, 10\degree,\, 20\degree]$ and $\vect{t}' = [5\degree,\, 15\degree,\, 20\degree]$ are not because they differ in the torsion angle value for $T_2$.

\subsection{Neighbourhood Search}
Based on the discussion in Section~\ref{subsec: binary formulation}, the problem of finding the best solution in $N_k(\vect{t})$ can be formulated as a QUBO problem by restricting the optimization problem in (\ref{eq: general binary optimization}) to the binary variables corresponding to the torsions in $\cal{T}_k$ and moving the one-hot encoding constraints to the objective function using the quadratic penalty method as follows:
\begin{equation}
\begin{aligned} \label{eq: neighbourhood qubo formulation}
\min & \sum_{\substack{i,j:T_i, T_j \in \cal{T}_k \\ i \neq j}} \sum_{k _i=1}^d \sum_{k_j=1}^d U_{ij}(\theta_{k_i},\theta_{k_j}) x_{ik_i} x_{jk_j}+ \sum_{i:T_i \in \cal{T}_k} \sum_{k_i = 1}^d U_i(\theta_{k_i}) x_{ik_i} + p \sum_{i:T_i \in \cal{T}_k} \left( \sum_{k_i=1}^d x_{i,k_i} -1 \right)^2, \\
\text{s.t.} &  \quad x_{ik_i} \in \{0,1\}, \quad \forall i, k_i\,. 
\end{aligned}
\end{equation}
Here, the $U_{ij}(\theta_{k_i},\theta_{k_j})$ terms represent the interaction energy of the two vertices of $G_k$ that are connected by $T_i$ and $T_j$ when $t_i=\theta_{k_i}$ and $t_j = \theta_{k_j}$. The $U_i(\theta_{k_i})$ terms represent the interaction energy of the two vertices connected by $T_i$ on $G_k$ when $t_i = \theta_{k_i}$ and $p$ is a sufficiently large penalty coefficient that enforces the one-hot encoding constraints. The above QUBO problem can be solved using various methods~\cite{kochenberger2014unconstrained, wang2012path, shi2017parallel}, as well as specialized hardware devices such as quantum annealers. 

\subsection{Neighbourhood Change}
At each iteration, the neighbourhood is selected based on a random ordering, $\phi:\cal{T} \rightarrow \cal{T}$, of the torsions. The pseudocode for the neighbourhood change function is given below.

\begin{figure}[h]
\begin{algorithm}[H]
\begin{algorithmic}
\State Generate a random ordering, $\phi=(\phi(1),\phi(2),\ldots,\phi(M))$, of $\cal{T}$ 
\State $\cal{T}' \leftarrow \emptyset$
\For{$i=1,\ldots,M$}
	\If{$\cal{T}' \cup \{\phi(i)\}$ is 2-torsion dependent}
		\State $\cal{T}' \leftarrow \cal{T}' \cup\{\phi(i)\}$ \Comment{Add $\phi(i)$ to $\cal{T}'$}        
    \EndIf
\EndFor
\\
\Return $N_k$ corresponding to $\cal{T}'$
\end{algorithmic}
\end{algorithm}
\caption{Neighbourhood change procedure.}
\label{Algo: neigbourhood change}
\end{figure}

\subsection{Practical Considerations}
Here, we detail the practical considerations of the proposed VND method to accommodate the use of existing and future quantum annealers.

The formulated QUBO problem in~\eqref{eq: neighbourhood qubo formulation} is fully connected, meaning that for all $i$ and $j$ ($i \neq j$), the term $x_{ik_i}x_{jk_j}$ appears in the objective function. On the other hand, the connectivity of the qubits on the D-Wave 2000Q follows a ``Chimera graph''; thus, the problem in~\eqref{eq: neighbourhood qubo formulation} must be embedded onto the hardware graph using an embedding strategy~\cite{biswas2017nasa}. There is a limit on the number of variables a fully connected QUBO problem that can be embedded onto the graph can have. We take this limitation into consideration by imposing a limit on the number of variables in the formulated QUBO problem for each selected subset of torsion angles $\cal{T}_k$. We denote this parameter by $s$. In the following, we explain how this limit is imposed on the formulated QUBO problem.

For the selected neighbourhood structure at each iteration, we (randomly) select a total of $s$ discrete values. That is, for each $T_i$ in $\cal{T}_k$, we randomly choose a set of $\Theta_i \subseteq \Theta$ discrete values such that $s = \sum_{i: T_i \in \cal{T}_k} \vert \Theta_i \vert .$ We have already defined the $N_k(\vect{t})$ neighbourhood  as the set of torsion angle vectors $\vect{t}'$ that are different from $\vect{t}$ only in the angle values of the torsions  in $\cal{T}_k$. In addition, for any $T_i$ in $\cal{T}_k$, $t'_i$ (i.e., the value associated with $T_i$ in $\vect{t}'$) takes on values only from $\Theta_i$. 

With the above choice of neighbours, for an arbitrary $\vect{t}$, $N_k(\vect{t})$ contains
\begin{equation} \label{eq: num solutions}
S_k = \prod_{i:T_i \in \cal{T}_k} \vert \Theta_i \vert
\end{equation}
solutions. However, to find the best solution in $N_k(\vect{t})$, we need to pre-evaluate only 
\begin{equation} \label{eq: num energy evals}
\sum_{\substack{i,j:T_i,T_j \in \cal{T}_k \\ i \neq j}} \vert \Theta_i \vert \vert \Theta_j \vert + s
\end{equation}
energy terms to formulate the QUBO problem. As seen above, $S_k$ grows linearly with the product of $\vert \Theta_i \vert$, for all $i$, whereas the growth of \eqref{eq: num energy evals} is quadratic.  This means that the number of energy pre-evaluations grows more slowly than the size of the neighbourhood as $\vert \cal{T}_k \vert$ increases. 

Another practical consideration for our proposed VND method is the stopping criteria. The first stopping criterion sets a limit on the computational effort of the method by introducing a maximum number of iterations, denoted by $B$. The second stopping criterion aims to terminate the method early if it becomes stuck at a local minimum or finds the global minimum of the problem. For this purpose, we introduce a parameter called $A$ that represents the maximum number of consecutive iterations to can be performed without decreasing the energy. 

With the above-mentioned practical considerations, the implemented VND method is summarized in Fig.~\ref{Algo: implemented VND}.

\begin{figure}[h]
\begin{algorithm}[H]
\begin{algorithmic}
\Require A random initial conformation of the molecule, denoted by a torsion vector $\vect{t}_{\mathrm{init}}$
\State $a , b \leftarrow 0$ \Comment{Initialize the counters}
\State Select an arbitrary $N_k\in \{N_1,\ldots, N_{K}\}$ \Comment{Initialize the neighbourhood structure}
\State $\vect{t}_{\mathrm{c}} \leftarrow \vect{t}_{\mathrm{init}}$ \Comment{Initialize the current solution $\vect{t}_\mathrm{c}$}
\While{$a < A$ and $b < B$}
    \State Formulate the QUBO problem to find the best solution in $N_k(\vect{t}_{\mathrm{c}})$
    \State $\vect{t'} \leftarrow $ arg$\min_{\vect{t} \in N_k(\vect{t}_{\mathrm{c}})} U(\vect{t})$ \Comment{Find the best solution in $N_k(\vect{t}_{\mathrm{c}})$}
    \If{$U(\vect{t}') < U(\vect{t}_{\mathrm{c}})$}
    	\State $\vect{t}_{\mathrm{c}} \leftarrow \vect{t}'$ \Comment{Update the current solution}
        \State $b  \leftarrow 0 $ \Comment{Reset the no-improvement counter}
    \Else
    	\State $b  \leftarrow b + 1$ \Comment{Update the no-improvement counter}
    \EndIf
    \State $a \leftarrow a + 1$ \Comment{Update the counter}
    \State $N_k \leftarrow $ NeighbourhoodChange() \Comment{Change neighbourhoods}
\EndWhile
\end{algorithmic}
\end{algorithm}
\caption{Conformational search VND method.}
\label{Algo: implemented VND}
\end{figure}

Two important features of the VND method is its scalability and ease of adaptation to the size and connectivity of the quantum annealer. More specifically, one can simply increase $s$ in the described method to take advantage of improvements in the number of qubits and their connectivity. The rest of the method remains intact. It is worth mentioning that although the focus here has been on quantum annealers, other QUBO problem solvers could be used.

\subsection{Effect of the Molecular Structure} \label{subsec: molecule structure}
As previously discussed, due to the limitations of current quantum annealers, we are able to jointly optimize only a subset of torsions that are 2-torsion dependent at each iteration of VND. The potential speedup of the quantum annealer over a na\"ive exhaustive QUBO problem solver depends on the cardinality of the selected subset of torsions. 

Since $\sum_{i: T_i \in \cal{T}_k} \vert \Theta_i \vert $ is fixed, the number of solutions in $N_k(\vect{t})$, $S_k$, is maximized when the cardinality of each $\Theta_i$ is the same. That is,  when $\vert \Theta_i \vert = s / \vert \mathcal{T} _k \vert$ for all $i \in \{1, \ldots, \vert \cal{T}_k \vert \}$, resulting in the maximum number of solutions
\begin{equation} \label{eq: max S_k}
S_k = \left( \frac{s}{\vert \mathcal{T}_k \vert} \right) ^{\vert \cal{T}_k \vert}.
\end{equation}
$S_k$ is an increasing function of $\vert \cal{T}_k \vert$ if $s / \vert \cal{T}_k \vert \geq e$, where $e$ is the base of the natural logarithm.

As the potential speedup of the quantum annealer over an exhaustive QUBO solver is dependent on $S_k$, it is favourable to select each $\cal{T}_k$ with a large cardinality. The speedup diminishes when $\vert \cal{T}_k \vert = 2$, which is the case for molecules with linear rigid-body graphs. The above discussion suggests that star-like molecules that have rigid-body graphs with high-degree nodes stand to benefit more (than those that do not) from using a quantum annealer to solve the QUBO problem at each VND iteration. 

\section{Experimental Results}\label{sec: results}
In this section, we evaluate the performance of the proposed VND method. We first provide details about the molecules used in our experiments and then present the experimental results. 

\subsection{Trial Molecules} 
\label{molecules}
\begin{figure}
\centering
\includegraphics[width=0.7\textwidth]{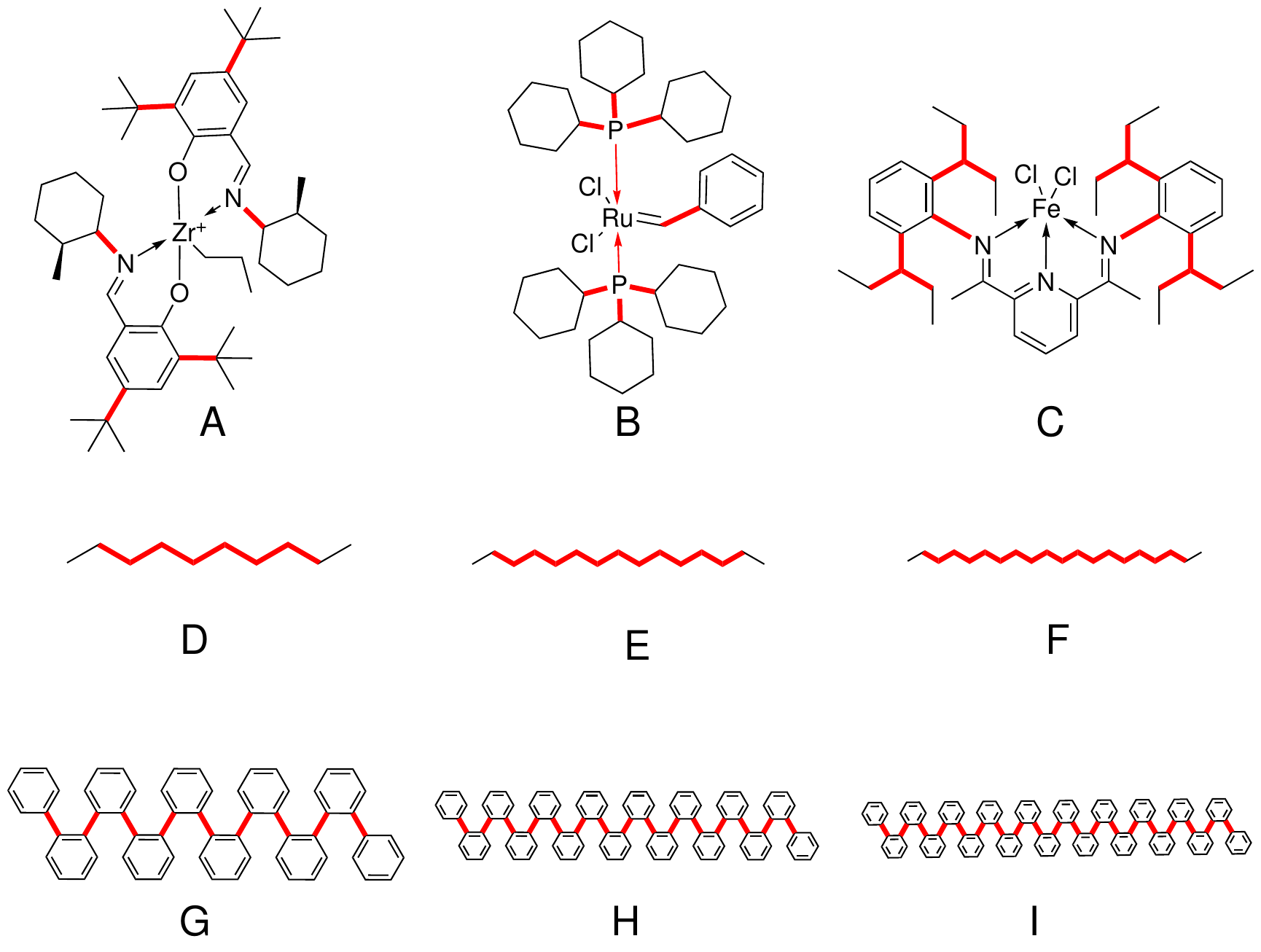}
\caption{Graphical representation of the six model systems studied: three organometallic compounds (A--C), three n-alkanes (D--F), and three ortho-phenylenes (G--I). The thick (red) lines represent the torsion bonds. See the Supplementary Information for additional details about each model system.}
\label{Fig: Model systems}
\end{figure}

The performance of VND was evaluated using a testbed containing the nine molecules depicted in Fig.~\ref{Fig: Model systems}. These model systems include three organometallic molecules useful for catalyzing reactions relevant to industry (labelled ``A''~\cite{Arriola2006,Makio2002}, ``B''~\cite{Torker2010,Schwab1995}, and ``C''~\cite{Kawakami2015,Small1998}), a set of three n-alkanes whose basic structural motif appears in fuels, lubricants, solvents \cite{ullmans}, and resins \cite{polyethylenehandbook} (labelled by D, E, and F), as well as a set of three ortho-phenylene oligomers that are of interest as electronic materials and nanomaterials  (labelled ``G'', ``H'', and ``I'')~\cite{He2010,Berresheim1999}.

The choice of model systems A--I was motivated by several considerations. First, it is important to show that our method can be applied to a wide variety of conformational search problems of relevance to industry. Second, the model systems are representative of a diversity of molecular graphs: systems A--C have star-like graphs, whereas D--I have linear graphs, albeit with different structures. Third, the model systems represent a significant variety of active torsions, ranging from very modest (e.g., A) to very substantial (e.g., F and I). Another motivation for choosing these systems was the existence of experimental data pertaining to their three-dimensional structure (see the Supplementary Information for details). 

\subsection{Energy Model}
In our experiments, the Lennard-Jones 6-12  potential is used to model the interaction energy of two atoms $\alpha$ and $\beta$ as
\begin{equation}
V(\alpha,\beta) = \epsilon_{\alpha\beta} \left[ \left(\frac {\sigma_{\alpha\beta}}{r_{\alpha\beta}} \right)^{12}  -2 \left(\frac {\sigma_{\alpha\beta}}{r_{\alpha\beta}} \right)^6 \right], 
\end{equation}
where $\epsilon_{\alpha\beta}$ is the depth of the potential well, $\sigma_{\alpha\beta}$ is the van der Waals bond length, and $r_{\alpha\beta}$ is the distance between the two atoms. An expression for equation \eqref{eq: U(t)} is then obtained by summing over all pairs of atoms. Values for the parameters $\epsilon_{\alpha\beta}$ and $\sigma_{\alpha\beta}$ are taken from the UFF~\cite{Rappe1992} without modification. One should note that while we use the Lennard-Jones potential in what follows to provide a proof of concept for the proposed VND method, other force fields, that allow for representing the molecular energy as equation~\eqref{eq: U(t)}, can also be used.

\subsection{Results}
We present the results of solving the conformational search  problem for the selected molecules using the proposed VND method. Since the D-Wave 2000Q quantum annealer is not guaranteed to find an optimal solution, we first present results for VND, where an exact QUBO problem solver is used. These results focus solely on the performance of the VND method as a conformational search approach by preventing any deterioration of the results attributable to the use of a device that has imperfections. For comparison purposes, we also present the results achieved by performing a local search heuristic~(LS), and a hybrid of the two methods (LS--VND) in which a random conformation is first optimized using LS and then passed to VND for further optimization. The comparison with LS is  helpful in understanding the improvement that can be achieved through the use of a more complex neighbourhood than what is used for LS. We also present the VND results with the quantum annealer used as the underlying QUBO problem solver. Finally, we compare the results with those found by our implementation of the PTMC algorithm for the conformational search problem, a state-of-the-art metaheuristic for conformational sampling. Details of the parameters used for VND, the quantum annealer, and PTMC are presented in the Supplementary Information.

\subsubsection{Reference Conformations} 
We compare our results against \emph{reference} conformations found using a PTMC conformational search method. More specifically, we use an {\em initial} conformation for each of the selected molecules as input to a PTMC algorithm and let it run with a sufficiently large number of sweeps, such that the resulting {\em reference} geometry can be assumed to represent the absolute minimum conformation with a high degree of confidence. Details of this procedure are given in the Supplementary Information. To ensure the fairness and accuracy of this comparison, the reference conformations are generated using the same potential energy model as the one we used for VND. 

\subsubsection{Performance Metrics}
The following metrics have been used for performance evaluation.
\begin{itemize}
\item Success rate: the fraction of runs that found a conformation with an energy within 1 kcal/mol (roughly chemical accuracy) of the reference conformation's energy.
\item Number of energy evaluations: the number of molecular energy evaluations needed to arrive at the best found conformation, averaged over all runs.
\item Residual: the energy difference between the best conformation found in a run and the reference conformation. The normalized residual, when reported, refers to the ratio of the residual to the number of atoms in the system.
\item Time to solution (TTS): the time it took to find the best solution (conformation) in a single run.	
\end{itemize}

Each run was terminated once it found a conformation within 0.1 kcal/mol of the reference conformation, or if it reached some other stopping criterion.

\subsubsection{VND vs. LS vs. LS--VND}
As a baseline for comparison, we used an exact QUBO problem solver to optimize the selected set of torsions at each VND iteration in runs of both VND and LS--VND. The success rate, residual, and TTS for each of the three methods, for all nine model systems, are presented in Table~\ref{table:ex_results}. \highlight{Here, the TTS for VND includes the time needed for selecting the neighbourhoods, formulating their associated QUBO problems, solving the QUBO problems with an exact solver, and translating the solutions of the QUBO problems back to torsion vectors. The number of energy evaluations reported for the VND method is based on the number of energy evaluations required to find the $U_{ij}(\theta_{k_i},\theta_{k_j})$ and $U_i(\theta_{k_i})$ coefficients of the QUBO problem in \eqref{eq: neighbourhood qubo formulation}.} As shown, whereas LS is faster than VND and LS--VND, its success rate and residual results are generally inferior to the other two methods. Our implementation of LS terminates when there is no neighbouring solution which has a lower energy value, ensuring that the algorithm terminates at a locally optimal solution with respect to the single-torsion neighbourhood. Further, since LS--VND combines both the LS and VND methods, it is not surprising that its success probability and residual are generally at least as good as those of LS and VND.

\begin{table}[h!]  
\centering
\begin{tabular}{llcc@{\hskip 0.4in}rrr@{\hskip 0.4in}rrr}
\hline
\hline
Method & Model & Success  & Num. energy & \multicolumn{3}{l}{\hskip -0.1in Normalized residual}     &  \multicolumn{3}{c}{TTS (seconds)}   \\
& system & rate & evaluations &  Min &   50th &   75th &   Min &    50th &    75th \\
\hline
\multirow{9}{*}{VND} 
    & A              &      1.00 & $4.6 \times 10^4$ &  0.00 &   0.00 &   0.00 &  19.7 &  90.2 & 110.2 \\
    & B              &      0.60 & $1.42 \times 10^5$ & 0.00 &   0.01 &   0.01 &  90.9 & 271.6 & 320.3 \\
    & C              &      0.04 & $2.11 \times 10^5$ & 0.00 &  0.13 &  0.16 & 127.3 & 418.7 & 498.3 \\
    & D              &      1.00 & $1.96 \times 10^4$ & 0.00 &   0.00 &  0.00 &   3.0 &  10.0 &  13.8 \\
    & E              &      1.00 & $4.22 \times 10^4$ & 0.00 &   0.00 &   0.00 &  11.7 &  25.9 &  30.7 \\
    & F              &      1.00 & $7.25 \times 10^4$ & 0.00 & 0.00 &   0.00 &  34.0 &  64.1 &  78.5 \\
    & G              &      0.16 & $1.23 \times 10^5$ & 0.00 &  0.14 &  0.21 &  37.0 & 101.1 & 153.5 \\
    & H              &      0.00 & $1.69 \times 10^5$ & 0.02 &  0.25 &  0.31 &  67.5 & 211.6 & 243.4 \\
    & I              &      0.00 & $1.87 \times 10^5$ & 0.01 &  0.30 &  0.38 & 155.7 & 374.1 & 391.4 \\
\hline 
\multirow{9}{*}{LS} 
    & A              &      1.00 & $1.45 \times 10^4$ &  0.00 & 0.00 &  0.00  &   7.8 &   9.6 &   10.6 \\
    & B              &      0.03 & $1.01 \times 10^5$ & 0.01 &  0.28 &  0.39 &  35.0 &  69.7 &   91.6 \\
    & C              &      0.02 & $2.47 \times 10^5$ & 0.00 &  0.16 &  0.21 &  84.4 & 162.4 &  198.9 \\
    & D              &      1.00 & $2.21 \times 10^4$ &  0.00 &   0.00 &   0.00 &   5.7 &   8.9 &    9.9 \\
    & E              &      1.00 & $7.17 \times 10^4$ &  0.00 &   0.00 &   0.00 &  26.5 &  38.4 &   41.8 \\
    & F              &      0.99 & $1.53 \times 10^5$ & 0.00 &   0.00 &   0.00 &  70.9 & 105.2 &  115.7 \\
    & G              &      0.03 & $1.87 \times 10^5$ & 0.00 &  0.15 &  0.18 &  36.3 & 112.9 &  148.2 \\
    & H              &      0.00 & $4.76 \times 10^5$ & 0.05 &  0.28 &  0.37 & 168.5 & 468.1 &  580.7 \\
    & I              &      0.00 & $8.10 \times 10^5$ & 0.09 &  0.29 &  0.39 & 354.6 & 784.2 & 1078.2 \\
\hline
\multirow{9}{*}{LS--VND} 
    & A              &      1.00 &  $1.45 \times 10^4$ & 0.00 &  0.00 &  0.00 &   7.5 &   10.3 &   11.4 \\
    & B              &      0.44 &  $2.23 \times 10^5$ & 0.00 &  0.01 &  0.13 &  68.9 &  294.5 &  421.7 \\
    & C              &      0.04 &  $3.58 \times 10^5$ & 0.00 &  0.12 &  0.16 & 135.6 &  305.8 &  474.3 \\
    & D              &      1.00 &  $2.16 \times 10^4$ & 0.00 &   0.00 &   0.00 &   5.7 &    8.7 &   10.0 \\
    & E              &      1.00 &  $7.19 \times 10^4$ & 0.00 &  0.00 &   0.00 &  25.8 &   38.1 &   40.9 \\
    & F              &      1.00 &  $1.48 \times 10^5$ & 0.00 &  0.00 &   0.00 &  76.6 &  101.0 &  109.3 \\
    & G              &      0.17 &  $2.30 \times 10^5$ & 0.00 &  0.15 &  0.18 &  26.3 &  142.0 &  199.0 \\
    & H              &      0.03 &  $5.88 \times 10^5$ & 0.00 &  0.23 &  0.31 & 272.6 &  576.6 &  742.7 \\
    & I              &      0.02 &  $8.42 \times 10^5$ & 0.00 &  0.30 &  0.39 & 435.1 & 1048.3 & 1284.8 \\
\hline
\hline
\end{tabular}
\caption{Results for VND, LS, and LS--VND, when using an exact solver as the underlying QUBO solver. For each model system we report the minimum, median, and 75th percentile of the residual and TTS, over 100 runs.}
\label{table:ex_results}
\end{table}

\subsubsection{VND Using a Quantum Annealer}
The results for VND using a quantum annealer as the underlying solver are presented in Table~\ref{table:bcd_dwave_results}. \highlight{Here, the TTS includes the time spent on all pre- and postprocessing steps needed to solve the QUBO problems on the quantum annealer. Note that the actual time spent using the quantum annealer, referred to as the annealing time, is much smaller than the reported TTS.} In addition, to aid in visually assessing the found lowest-energy conformations, their graphical representations are provided in the Supplementary Information. A comparison of these results with the results of VND when using an exact QUBO problem solver (see Table~\ref{table:ex_results}) shows that those of the former are of lower quality. It is expected, however, that spending more effort on tuning the quantum annealer's parameters would improve these results (See Supplementary Information for more details). Another observation from Table~\ref{table:bcd_dwave_results} is that VND employing a quantum annealer to solve QUBO problems has a larger TTS than VND using an exact QUBO problem solver. The quantum annealer solves a QUBO problem much faster than the exact solver, so one might think that its TTS should also be lower.

To explain this observation, one should note that it is not sufficient to solve QUBO problems merely fast, because if they are not solved optimally by the quantum annealer, the VND method may take longer to converge. Further, a significant portion of time is spent on transforming the QUBO problems into Ising problems (see the Supplementary Information), communicating with the quantum annealer, and mapping the results from the quantum annealer back to the logical bits. It is worth noting that elapsed real time is a fair measure of the time required to solve actual problems using a quantum annealer. This is in contrast to the customary approach of reporting only the annealing time, which is very small in comparison (on the order of a few microseconds per annealing cycle).   

There are several known factors that make the QUBO problems generated at the VND iterations challenging for the current generation of quantum annealers. First, the problem graphs are fully connected, whereas the quantum annealer's connectivity graph is extremely sparse, resulting in each logical bit being embedded onto chains of 17 qubits on the quantum annealer. It is difficult to maintain identical states for those qubits, resulting in a higher error rate. Second, these problems have a very large range of coefficients, due to the $1/r^{12}$ and $1/r^6$ terms in the Lennard-Jones 6-12 energy model. On the other hand, the couplers of the existing quantum annealers have a limited bit precision and a fixed range. The large range of coefficients results in a loss of precision, which manifests itself in a lower success probability. Third, the current generation of quantum annealers has a high level of noise, referred to as intrinsic control error (ICE), leading to a significant loss in precision. Future quantum annealers are expected to mitigate these factors, with more-dense hardware graphs, higher bit precision, and lower ICE levels.

\begin{table}[H]  
\centering
\begin{tabular}{lcc@{\hskip 0.4in}rrr@{\hskip 0.4in}rrr}
\hline
\hline
Model & Success  & Num. energy & \multicolumn{3}{l}{\hskip -0.1in Normalized residual}     &  \multicolumn{3}{c}{TTS (seconds)}   \\
system & rate & evaluations &  Min &   50th &   75th &   Min &    50th &    75th \\
\hline
 A        &            0.00 &  $4.56 \times 10^4$ &  0.01 & 0.05 & 0.06 & 156.1 &  330.2 &  366.4 \\
 B        &            0.00 &  $4.99 \times 10^4$ &  0.12 & 0.47 & 0.76 & 136.0 &  288.9 &  395.5 \\
 C        &            0.00 &  $9.03 \times 10^4$ &  0.14 & 0.38 & 0.62 & 300.9 &  507.3 &  726.9 \\
 D        &            1.00 &  $3.86 \times 10^4$ & 0.00 &  0.00 & 0.01 & 299.0 &  516.4 &  608.0 \\
 E        &            0.72 &  $8.24 \times 10^4$ & 0.00 &  0.01 & 0.02 & 235.8 &  639.1 &  815.6 \\
 F        &            0.60 &  $1.17 \times 10^5$ & 0.00 &  0.02 & 0.02 & 533.7 &  961.4 & 1038.8 \\
 G        &            0.00 &  $9.28 \times 10^4$ & 0.08 &  0.24 & 0.34 & 385.8 &  788.3 &  990.1 \\
 H        &            0.00 &  $1.27 \times 10^5$ &  0.19 & 0.43 & 0.64 & 445.7 &  982.2 & 1107.3 \\
 I        &            0.00 &  $1.7 \times 10^5$ &  0.25 & 0.52 & 0.75 & 511.1 & 1221.2 & 1648.8 \\
\hline
\hline
\end{tabular}
\caption{Results for VND using a quantum annealer as the underlying QUBO solver. For each model system we report the minimum, median, and 75th percentile of the residual and TTS, over 25 runs.}
\label{table:bcd_dwave_results}
\end{table}

\subsubsection{Comparison with PTMC}
The results for PTMC are presented in Table~\ref{table:pt_results}. PTMC had a high success rate for all model systems except for system F. It is worth noting that PTMC's TTS and number of energy evaluations are generally significantly higher than that of VND.

\begin{table}[H]  
\centering
\begin{tabular}{lcc@{\hskip 0.4in}rrr@{\hskip 0.4in}rrr}
\hline
\hline
Model & Success  & Num. energy & \multicolumn{3}{l}{\hskip -0.1in Normalized residual}     &  \multicolumn{3}{c}{TTS (seconds)}   \\
system & rate & evaluations &  Min &   50th &   75th &   Min &    50th &    75th \\
\hline
 A              &      1.00 & $1.12 \times 10^5$ &  0.00 &  0.00 & 0.00 &   40.1 &  107.5 &  143.3 \\
 B              &      1.00 & $3.83 \times 10^6$ &  0.00 &  0.00 & 0.00 &  301.0 &  787.5 & 1092.8 \\
 C              &      1.00 & $3.19 \times 10^6$ &  0.00 & 0.00 & 0.00 &  585.5 & 2068.3 & 3025.0 \\
 D              &      1.00 & $4.05 \times 10^4$ & 0.00 &   0.00 & 0.00 &   52.1 &   87.9 &  108.0 \\
 E              &      1.00 & $1.03 \times 10^5$ &  0.00 &   0.00 &   0.00 &  195.6 &  290.5 &  326.3 \\
 F              &      1.00 & $2.00 \times 10^5$ &  0.00 &   0.00 &   0.00 &  457.4 &  685.3 &  776.0 \\
 G              &      1.00 & $2.52 \times 10^6$ &  0.00 &   0.00 &   0.00 &  154.8 &  710.4 & 1085.0 \\
 H              &      0.60 & $5.54 \times 10^6$ &  0.00 &   0.00 &   0.05 & 1051.4 & 4711.6 & 5975.8 \\
 I              &      0.12 & $3.58 \times 10^6$ &  0.00 &  0.07 &  0.09 & 3025.9 & 5758.3 & 6631.8 \\
\hline
\hline
\end{tabular}
\caption{Results for PTMC. For each model system, we report the minimum, median, and 75th percentile of the residual and TTS, over 100 runs.}
\label{table:pt_results}
\end{table}

\subsubsection{Effect of the Neighbourhood Size on VND's Performance}
In the VND experiments discussed above, we restricted the neighbourhood size, $s$, in each VND iteration to 63, the largest size of QUBO problem, whose underlying graph is complete, that can be solved using an equal-length embedding on the quantum annealer. In what follows, we report on the effect of $s$ on the performance of VND. This is useful in predicting the performance improvement achievable by increasing either the number of qubits or their connectivity.

Table~\ref{table:neighbourhood_size} presents the results for molecules B and C for different neighbourhood sizes when an exact solver is used to solve the QUBO problem at each iteration. The reason for choosing these two molecules is that they have a star-like structure and, as discussed in Section~\ref{subsec: molecule structure}, the advantage of using a quantum annealer over an exact solver is expected to be more pronounced for these molecules. 

As shown, increasing the neighbourhood size from 30 to 60 and then from 60 to 90 noticeably improves the results for both molecules. However, the improvement exhibits diminishing returns when the neighbourhood size is increased beyond 90. We expect similar behaviour to occur at different neighbourhood sizes for different families of molecules. 

\begin{table}[H]  
\centering
\begin{tabular}{clc@{\hskip 0.4in}ccc}
\hline
\hline
Neighbourhood & Model  & Success  &  \multicolumn{3}{c}{Residual (kcal/mol)} \\
Size ($s$) & System & Rate  &   Min &   50th &   75th \\
\hline
\multirow{2}{*}{30}  & B   &      0.07 &   0.2 &   16.7 &   26.0  \\
                     & C   &      0.00 &   1.1 &   19.4 &   25.6 \\ \hline
\multirow{2}{*}{60}  & B   &      0.39 &   0.0 &   1.3 &   1.9 \\
 					 & C   &      0.04 &   0.1 &   13.4 &  17.1 \\ \hline
\multirow{2}{*}{90}  & B   &      0.55 &   0.0 &   0.9 &   1.5 \\
 					 & C   &      0.09 &   0.0 &   8.7 &   14.8 \\ \hline
\multirow{2}{*}{120} & B   &      0.58 &   0.0 &   0.9 &   1.5 \\
 					 & C   &      0.10 &   0.1 &   8.7 &   14.1 \\
\hline
\hline
\end{tabular}
\caption{Effect of the neighbourhood size on the performance of VND. For each model system, we report the minimum, median, and 75th percentile of the residual, over 500 runs.}
\label{table:neighbourhood_size}
\end{table}
\section{Conclusion}\label{sec: conclusion}
In this paper, we have presented a variable neighbourhood descent (VND) method for conformational search. We introduced the concept of a rigid-body graph and used this simplified molecular structure to carefully define a neighbourhood structure to allow for efficient optimization using a binary quadratic optimizer. Based on current quantum annealing hardware, we selected a 2-torsion-dependent neighbourhood at each iteration such that finding the best solution in the selected neighbourhood could then be formulated as a QUBO problem. The size of the neighbourhood can be chosen such that the method can be adapted to the number of available qubits as well as to their connectivity on the quantum annealer. As a result, the proposed method is not only well-suited for current hardware, but can easily be adapted to take advantage of hardware improvements. Whereas the proposed method can be used as a standalone conformational search approach, it can be combined with existing conformational search methods for potentially improved performance. 

Beyond a simple presentation of the method, we also conducted a preliminary case study based on an implementation of the VND method using the D-Wave 2000Q quantum annealer for two families of molecules. In this exploration, we compared the results of our method with those of PTMC, a state-of-the-art solver for conformational search. To understand how much of the gap between the results of PTMC and those of VND used with a quantum annealer can be attributed to the imperfections of the quantum annealer, we replaced it with an exact QUBO problem solver. VND used along with the exact solver was able to find noticeably better conformations than those found using VND and the quantum annealer together. This observation points to the potential improvement achievable in the short term through more-advanced tuning of the existing quantum annealer, and in the long term using improved hardware. 

This work suggests a number of possible future research directions. For example, investigating refined neighbourhood change functions rather than using a random ordering of the torsion angles to choose the neighbourhood could lead to significant improvements. This could involve further exploitation of the molecular graph or the solutions from previous VND iterations. In addition, an improved selection of the $s$ discrete points at each VND iteration to account for hardware limitations could further improve the results. We leave these improvements for future work. 

We believe that the proposed method, based on careful hardware-aware neighbourhood selection, holds the potential to provide promising solutions to important optimization problems. \highlight{While PTMC shows better performance over the  molecules studied, our method has opened a scalable path forward for leveraging emerging quantum technologies for conformational search, a critically important problem in the field of chemical and materials science.}

\section{Acknowledgements}\label{sec: acknowledgements}
The authors thank Jamie Cohen, Alejandro Garza, Steve Arturo, Maritza Hernandez, Helmut G. Katzgraber, Kausar N. Samli, Takeshi Yamazaki, and Arman Zaribafiyan for their insightful comments, Varinia Bernales for assistance in preparing molecular geometries, and Marko Bucyk for reviewing and editing the manuscript. 

\bibliography{conf_search.bib}
\bibliographystyle{unsrt}


\section*{Appendix}\label{sec: appendix}

\subsection*{A Brief Overview of Quantum Annealing} \label{subsec: QA intro}
The D-Wave 2000Q quantum annealer solves problems in the form of an Ising Hamiltonian defined by
\begin{align*}
\displaystyle{\argmin_{\vect{s}}} \quad & \vect{s}^T \vect{J} \vect{s} + \vect{h}^T \vect{s}, \\
\mbox{s.t.} \quad &\vect{s} \in \{-1, 1\}^N,
\end{align*}
where $N$ is the number of qubits and $\vect{h}$ and $\vect{J}$ represent the local fields and couplers, respectively. The above Ising problem can be transformed into a quadratic unconstrained binary optimization (QUBO) problem
\begin{align*}
\displaystyle{\min_{\vect{x}}}  \quad & \vect{x}^T \vect{Q} \vect{x},   \\
\mbox{s.t.} \quad & \vect{x} \in \{0, 1\}^N
\end{align*}
by substituting $\vect{s} = 2 \vect{x}- \vect{1}$. To utilize the quantum annealer, the local fields and the pairwise couplings of the Ising formulation ($\vect{h}$ and $\vect{J}$) must be specified.

Quantum annealing is inspired by the adiabatic principle: a system at equilibrium, in its ground state, will remain in the ground state provided the system is evolved sufficiently slowly. During each annealing cycle, the Hamiltonian of the system is continuously deformed from a known, simple Hamiltonian $H_{0}$ to the problem Hamiltonian $H_{ \text{Ising}}$, which is typically a non-trivial task. This evolution is represented by the Hamiltonian

\begin{equation}
H(\tau) = A(\tau) H_{0}+ B(\tau) H_{ \text{Ising}}, \quad \tau \in [0,1]\,,
\end{equation}
where $\tau = t / t_{a}$, with $t_{a}$ representing the annealing time. The functions $A(\tau)$ and $B(\tau)$ specify the annealing schedule. Typically, $A(\tau)$ and $B(\tau)$ are monotonically decreasing and increasing functions, respectively. 

Given that the quantum annealer operates at a very low but finite temperature, the quantum annealer is a heuristic solver. For this reason, typical use of the quantum annealer involves obtaining an ensemble that consists of the results of hundreds, if not thousands, of reads from the quantum annealer. Additional specifications of the quantum annealer include the annealing time, the scaling of the $\vect{h}$ and $\vect{J}$ values, and using gauges to compensate for errors due to noise. We refer the interested reader to \cite{biswas2017nasa} for more details on quantum annealers and their applications.

\subsection*{Choice of Molecules}

\subsubsection*{Source (Experimental) Structures}
For systems A--C in Fig. 4 of the paper, atomic coordinates from single-crystal X-ray structures are available, either for the compounds themselves (B and C) \cite{groom2016,Torker2010,ccdc1416774} or for close analogues (A) \cite{Terao2006}. Systems A and C represent the active, cationic (A) and inactive, initial form (C) of two olefin polymerization catalysts. System B embodies a catalyst for olefin metathesis. \mbox{X-ray} experimental data pertaining to the three-dimensional (3D) structures of ortho-phenylenes (oPh) are currently confined to substituted analogues up to chain lengths of ten phenylene units \cite{Mathew2014}; the longest-chain unsubstituted oPh$^{x}$ instance is for $x=5$ \cite{ccdcgomessantoslow,ccdc808554,groom2016}. Complementary computational studies exist for $x=12$ and support the experimental 3D structures \cite{He2010}. Experiments and computation agree that the preferred morphology for oPh$^{x}$ oligomers is the tightly helical, ``closed-helix'' conformation. In the following, we refer to the geometries that correspond directly to experiments (or high-level computation for system A) as {\em source} geometries. 

\subsubsection*{Initial Conformations}
In the present work, we have focused exclusively on the torsional degrees of freedom, while other molecular degrees of freedom such as the bond lengths and angles were fixed at the values they assumed in the {\em initial} structure and excluded from the optimization. We employed two strategies to create initial structures. The details of these methods are given below; the geometries resulting from them are referred to as initial geometries in the following.
\begin{itemize}
{\item {\bf Method a:}}
Guesses pertaining to the structure of metal compounds were created using off-the-shelf molecular modelling software  based on the structure of experimentally characterized analogues (compound A) \cite{Terao2006,ccdc635962,ccdc635963} or directly obtained from the X-ray diffraction structures (compounds B and C) \cite{groom2016,ccdc811503,ccdc1416774}. These structures were locally optimized using the ORCA 4 program \cite{NeeseORCA,orcafaccts} in order to remove minor inconsistencies in the atomic positions. For this, we used the popular M06 density functional \cite{Zhao2008}  and a def2-SVP basis set~\cite{Weigend2005}, along with the corresponding fit basis set on non-metal atoms and a def2-TZVP basis set~\cite{Weigend2006} on metal atoms, together with the appropriate ECPs \cite{Andrae1990}. The RI-JK \cite{Weigend2002}  and RIJCOSX \cite{Neese2009} approximations were employed for acceleration. An unrestricted wave function with four unpaired spins was used for system C. 
{\item {\bf Method b:} }
Ortho-phenylene (oPh) geometries were generated as closed helices \cite{He2010} using Materials Studio \cite{bioviams} by optimizing them with the Universal force field \cite{Rappe1992} implemented in Materials Studio. There is evidence that the open and closed helical states of oPh$^{n}$ are close in energy \cite{He2010} to the majority of experimental evidence gathered in the solid state pointing to a general preference for the closed helix \cite{Ohta2011,Blake1998,Mathew2013,Mathew2014,ccdc954742}.  Using UFF to optimize all bond lengths, angles, and torsions, we found that the closed and open helix conformers of oPh$^{20}$ differ by 1.5 kcal/mol per phenylene unit, favouring the closed helix. \par
\end{itemize}

\subsubsection*{From Initial Conformations to Reference Conformations}
{\em Reference} conformations for all nine molecules, embodying a best-effort guess of the global minima available to the {\em initial} conformations in the torsional space, were generated using an implementation of the parallel tempering Monte Carlo (PTMC) method. All other experiments described in the paper were stopped once a conformation with an energy within 0.1 kcal/mol of the energy of the reference conformation was found.

The agreement between {\em source} structures, which closely mirror or are identical to experimental structures, and reference structures found by PTMC is significant for several reasons. Source structures embody the best available information regarding the global minimum geometry of the system. In order to be useful, a conformational search algorithm should be able to locate this geometry starting without the use of prior knowledge of the optimal set of torsion angles. The reference structure, therefore, should closely approximate the source structure. If they do not agree, several causes may be contributing to the disagreement:
\begin{itemize}
{\item {\bf Force field failure:}} The force field guiding the search of the torsional energy surface may be insufficiently realistic to capture the true global minimum.
{\item {\bf Structural bias:}} The initial set of bond lengths and bond angles supplied with the initial structure may bias the conformational search in the torsional space toward regions inconsistent with the  source geometry.
{\item {\bf Experimental bias:}} The source geometry itself may be dominated by effects not present in an isolated, molecular system, as might be the case if the source geometry is taken from a crystal with significant intermolecular interactions.
\end{itemize}
Table \ref{table:model_systems} summarizes the differences between  source and reference geometries. 
Agreement between the reference structures and the source structures is indicated by the contents of the ``Reference Matches Source" column. In general, reference structures correspond  closely to the source structures; except for the case of system B, source and reference structures occupy the same energy basin. For system B, the PTMC search identified a basin with a minimum energy lower than the source structure by $\approx 1.5$ kcal/mol. This deviation is attributable either to force-field failure (due to the use of a simplified Universal force field (UFF) during PTMC) or experimental bias (see above) due to subtle effects of crystal packing on the source geometry.

\begin{table}[H]  
\centering
\begin{tabularx}{0.7\textwidth}{XlrllX}
\hline
Model \newline system & Type & Torsions & Source & Method &	Reference matches source\\
\hline

A & metal comp'd & 6 & TW & a & Y\\
B &  metal comp'd & 9 & \cite{groom2016,Torker2010} & a &N\\
C &  metal comp'd & 14 & \cite{groom2016,ccdc1416774} & a &Y\\
D & n-alkane & 7 &  \cite{Luttschwager2013} & b &Y\\
E & n-alkane & 12 & \cite{Luttschwager2013} & b &Y\\
F & n-alkane & 17  & \cite{Luttschwager2013} & b & N\\
G & o-phenylene & 9 & TW & b &Y\\
H & o-phenylene & 15 & TW & b &Y\\
I & o-phenylene & 19 & TW & b &Y\\
\hline
\hline

\end{tabularx}
\caption{Geometric properties of model systems A--I. ``Model system'' gives the label assigned to each model, ``Type'' is the type of the model system, ``Torsions'' is the number of free rotating torsion bonds, ``Source'' is the source for the initial conformation (where ``TW'' stands for ``this work''), ``Method'' is the method for optimizing the source conformation to yield the initial conformation, and ``Reference matches source" indicates whether the source and reference conformations occupy the same energy basin.}
\label{table:model_systems}
\end{table}

\subsection*{Lowest-Energy Conformations Found by VND}
\begin{figure}[h!]
    \centering
    \includegraphics[width=0.5\textwidth]{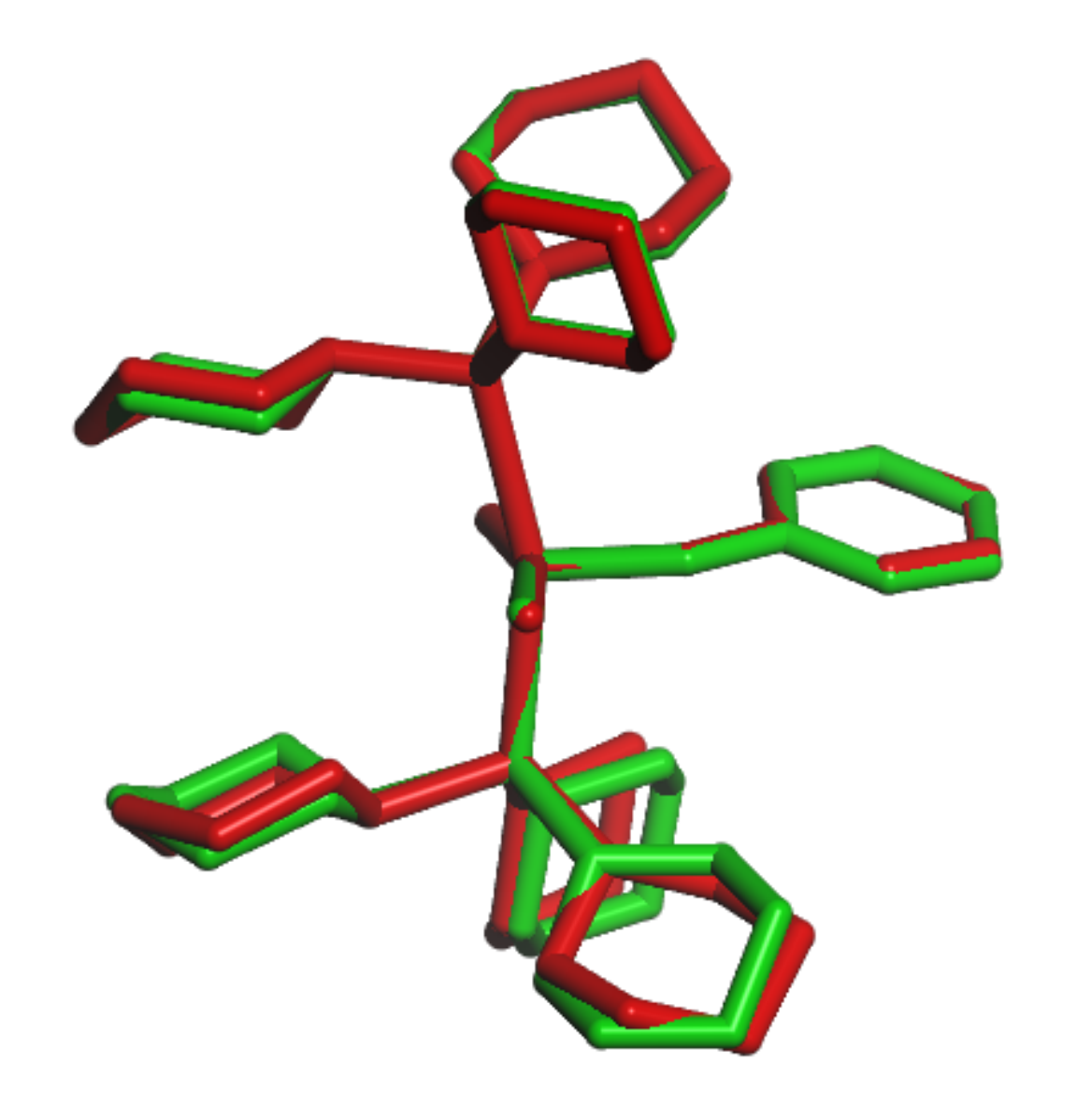}
    \caption{Overlay of the reference conformation (green) and the best conformation found by the VND method using a quantum annealer (red) for model system B.}
    \label{fig: IKORIK optimized}
\end{figure}

\begin{figure}[h!]
    \centering
    \includegraphics[width=0.7\textwidth]{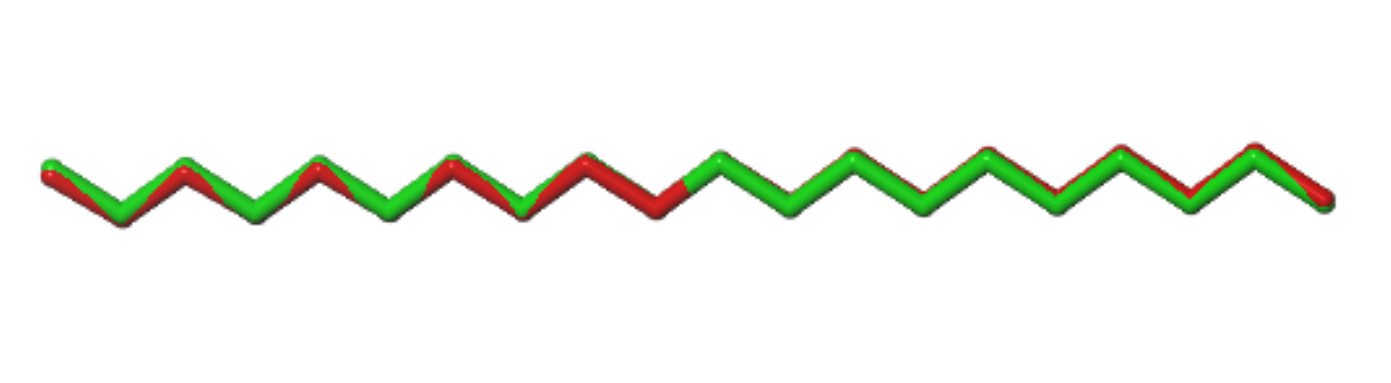}
    \caption{Overlay of the reference conformation (green) and the best conformation found by the VND method using a quantum annealer (red) for model system F.}
    \label{fig: eicosane optimized}
\end{figure}

\begin{figure}[h!]
    \centering
    \includegraphics[width=0.7\textwidth]{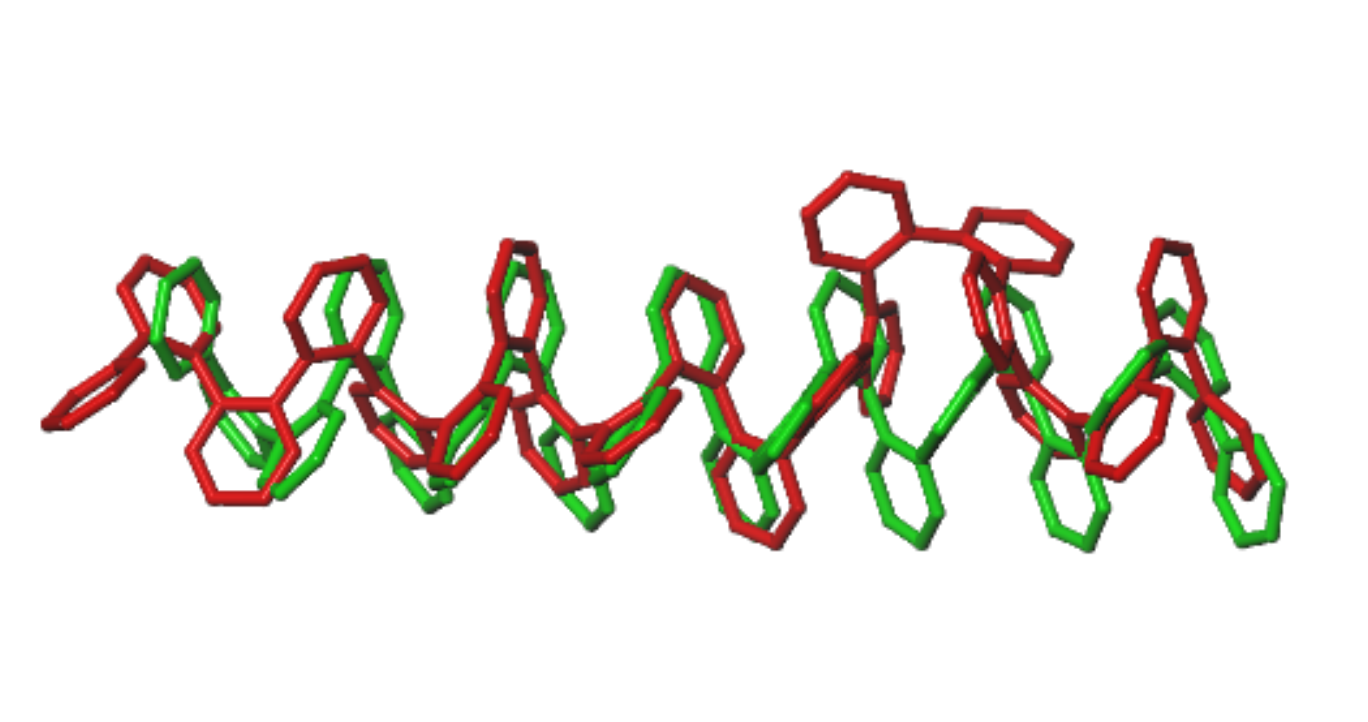}
    \caption{Overlay of the reference conformation (green) and the best conformation found by the VND method using a quantum annealer (red) for model system I.}
    \label{fig: ohiph10 optimized}
\end{figure}

Table 1 and Table 2 of the Results section of the paper present energy residuals of the conformers produced by VND. To provide a better understanding of the quality of the conformers, Table \ref{tab:my_table_rms} presents the root mean square (RMS) distance between the torsion vector of the reference conformer and the best conformer found by both VND used with the quantum annealer and VND used with an exact solver. The latter method provides an estimate of the potential gains of  improvements to the hardware. We see that in this case the best conformers are very close to the reference conformers and clearly in the same basin. On the other hand, the results of VND when using the quantum annealer have larger RMS distances. One reason there are larger RMS distances when using the quantum annealer even for small residuals is symmetries. Taking model system A, for example, there are conformations of the tertiary butyl group that are rotated by 60 degrees, a change that does not have a large impact on the energy. Due to these symmetries, RMS distances are of limited usefulness when comparing molecular geometries. To address this limitation, we present graphical representations of the lowest-energy conformations found for model systems B, F, and I shown as an overlay on the reference conformations (see Fig.~\ref{fig: IKORIK optimized}, Fig.~\ref{fig: eicosane optimized}, and Fig.~\ref{fig: ohiph10 optimized}). We also provide a file with all initial, reference, and optimized molecular conformations.   

    \begin{table}[h]
        \centering
        \begin{tabular}{|c|c|c|}
             
             \hline
             Model system & VND and quantum annealer & VND and exact solver \\ \hline
             A & 73.12$^\circ$ & 0.41$^\circ$ \\ \hline
             B & 10.54$^\circ$ & 0.47$^\circ$ \\ \hline
             C & 89.24$^\circ$ & 1.22$^\circ$ \\ \hline
             D & 2.29$^\circ$ & 0.91$^\circ$ \\ \hline
             E & 1.46$^\circ$ & 1.07$^\circ$ \\ \hline
             F & 2.06$^\circ$ & 0.00$^\circ$ \\ \hline
             G & 78.20$^\circ$ & 2.03$^\circ$ \\ \hline
             H & 114.86$^\circ$ & 5.96$^\circ$ \\ \hline
             I & 84.56$^\circ$ & 3.96$^\circ$ \\ \hline
        \end{tabular}
        
        \caption{Root mean square distance between the torsion vector of the reference conformation and the best conformation found with the VND method using the quantum annealer and the VND method using exact solver.}
        \label{tab:my_table_rms}
    \end{table}

\subsection*{Further Details of the Experiments}
Parameters of different methods for generating the experimental results are explained in the following.
\begin{itemize}
\item \textbf{PTMC}: Our implementation of PTMC performs $10$ parallel MC experiments (replicas) at different temperatures. Each sweep of PTMC consists of proposing $M$ moves, one for each torsion, for each of the replicas (i.e., a total of $10 M$ proposals to sweep over all replicas). The proposed moves are accepted according to the Metropolis criterion. Initial temperatures of the replicas are distributed geometrically, given the temperatures of the first and last replicas. The temperatures are automatically updated during the experiment to maintain reasonable Monte Carlo and replica exchange acceptance probabilities.
\item \textbf{D-Wave 2000Q}: To solve each of the QUBO problems created for optimizing a subset of torsions, a total of $5000$ solutions (reads) from the the quantum annealer are obtained and the best solution from among them is used. To improve the results obtained from the annealer, $10$ gauges are applied, meaning that the $5000$ solutions are obtained by making $10$ calls, each with a different gauge, to the quantum annealer, where each call returns $500$ solutions. The annealing time is $5$ microseconds for each of the reads. Using a deterministic embedding solver that finds complete embeddings with chains of equal length~\cite{boothby2016fast}, the largest complete graph that can be embedded on the quantum annealer has a size of $63$. Majority voting is used to settle the disagreement between the qubits of the chain assigned to a logical binary variable. 
\item \textbf{VND}: The VND results are obtained for $b_{\mathrm{max}} = 200$ and $b_{\mathrm{nc}}=10$. To respect the size of the largest  complete graph embeddable onto the quantum annealer's hardware graph, we set $s = 63$.
\end{itemize}

\subsection*{Notes on Quantum Annealer Parameter Tuning}

There are several parameters affecting the performance of the D-Wave 2000Q quantum annealer in its solving of QUBO problems. Some, such as the connectivity of the qubits and the intrinsic control error (ICE), are the specifications of the hardware over which the user does not have control, while other parameters are tunable. Among those that are tunable, we focused on the annealing time, the effect of having different numbers of gauges, the number of reads (i.e., the solutions obtained from the quantum annealer for the same Ising problem), and the strength of the couplings between the qubits in a chain assigned to a logical bit (here called the  \emph{scale}). Further, we explored the effect of the penalty value used for incorporating the one-hot encoding constraints on the quality of solutions. 

To tune the above parameters, we generated a small number of QUBO problems for each molecule, created at different iterations of the VND method, and used them as test cases for evaluating the parameter settings. To assess the performance of each set of parameters, we considered the best solution and the percentage of solutions that were feasible (i.e., those that satisfy all one-hot encoding constraints) within a given number of reads from the quantum annealer. Our tuning effort resulted in the choice of parameter settings introduced in the previous section. 

We note that the selected set of families of molecules have distinct properties based on the QUBO coefficients and the connectivity of the variables generated during iterations of VND. Our parameter tuning strategy may result in obtaining a set of parameters that perform relatively well for all families of molecules taken together, but may not necessarily be the best for each of them separately. Thus, we expect that a family-based parameter tuning could yield improved results. This can be achieved by performing a greater number of tuning iterations, or by considering a broader set of tunable quantum annealer parameters, both requiring additional quantum annealing time. 


\end{document}